# Alignment of electron optical beam shaping elements using a convolutional neural network

E. Rotunno[1], A.H. Tavabi[2], P. Rosi[3], S. Frabboni[1,3], P. Tiemeijer[4], R.E. Dunin-Borkowski[2], V. Grillo[1]

1) Istituto di Nanoscienze - CNR, 41125 Modena, Italy
2) Ernst Ruska-Centre for Microscopy and Spectroscopy with Electrons, Forschungszentrum Jülich, 52425 Jülich, Germany
3) University of Modena and Reggio Emilia, via G. Campi 213/A, 41125 Modena, Italy
4) Thermo Fisher Scientific, PO Box 80066, 5600 KA Eindhoven, The Netherlands

**Abstract**

A convolutional neural network is used to align an orbital angular momentum sorter in a transmission electron microscope. The method is demonstrated using simulations and experimentally. As a result of its accuracy and speed, it offers the possibility of real-time tuning of other electron optical devices and electron beam shaping configurations.

**Introduction**

The evolution of resolution and optics in electron microscopy has involved major steps, including the introduction of spherical aberration correction [1,2], chromatic aberration correction [3,4] and monochromators [5,6,7,8]. Each increase in complexity has resulted in an increase in the difficulty of instrument control. Although the basic concept of the operation of each lens and optical element is known, the overall behaviour of the microscope is not predictable in detail and the quality of microscope performance is limited by the skill of the user.

One of the most significant recent developments in electron microscopy is electron beam shaping through the use of material-based holograms [9,10,11,12] and, more recently, electron optical components based on microelectromechanical systems technology [13,14,15]. Electron beam shaping can be used to generate vortex beams [16,17], non-diffracting beams [18], compact aberration correctors [19,20,21] and analysers of quantum states by means of unitary wave transformations [22]. These ideas are often inspired by light optics, where their implementation is easier. Both standard electron beam control and new ideas of electron beam shaping require more automated control of the electron column, both to increase the speed and reproducibility of electron optical alignment and to reduce the demand on the operator.

In visible light optics, every element can be positioned manually and aligned separately, with adaptive optics providing improvements in telescope optics [23]. In electron microscopy, hardware aberration correctors formed from high order multipoles permit the limitations of cylindrically symmetric lenses to be circumvented. They are presently controlled using semi-analytical models [24,25,26,27] based on parametrized aberrations and on the effect of each multipole excitation. However, such an approach cannot easily be adapted to more unconventional optics, such as electron beam shaping. Moreover, a general approach is required to control a full microscope.

Here, we use a convolutional neural network (CNN) [28,29], which is now implemented in many other scientific disciplines [30,31,32,33,34] and is able to learn from a large set of training images to extrapolate a detail or the value of a parameter tagged to each image [35,36]. The success of the technique is based on the fact that it permits any parametrical space to be treated, no matter how complex, provided that enough data are fed to the learning algorithm. It is also often more "robust" to noise than an analytical model.

Here, we are motivated by the specific case of an orbital angular momentum (OAM) sorter [37,38,39], which makes use of electron beam shaping to measure an electron beam's component of OAM in the propagation direction by decoupling the azimuthal and radial degrees of freedom. Apart from diffraction, it provides the first complete example of a lossless unitary base change that "diagonalizes" a quantum operator using wave manipulation. Recent research [40] suggests that it could be the first of many useful wave transformations to revolutionize the concept of measurement in electron microscopy. The implementation of an OAM sorter requires precise alignment and control of two optical phase elements. No simple analytical model can be used to completely predict the effect on a final OAM spectrum of the available control parameters. In this paper, we perform a quantitative comparison with experiments to show the use of a CNN to control the alignment parameters of an OAM sorter.

**Methods**

- **Sorter misalignment**

An OAM sorter comprises two phase elements, which can each be based on synthetic holography or electrostatic potentials and are used to spatially separate the different OAM components of an electron beam. In the stationary phase approximation, the first element imparts a coordinate transformation from Cartesian to log-polar coordinates to an electron wave. The transformation phase is removed by exact phase compensation of second phase elements positioned in the Fraunhofer plane of the first element. The working principle is shown schematically in Figure 1.

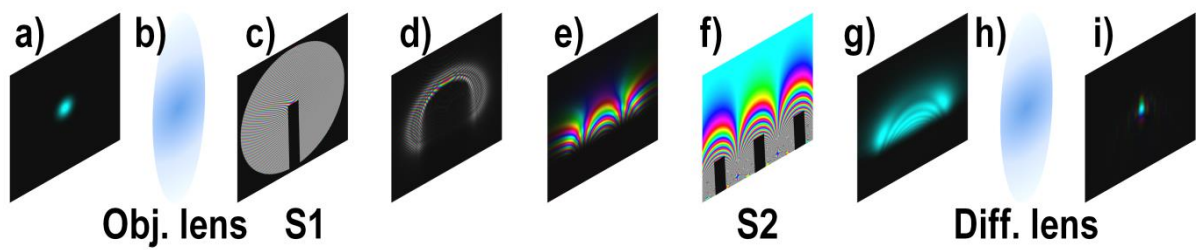

Figure 1: Schematic diagram showing the evolution of an electron beam through an orbital angular momentum sorter. See text for details.

In the most natural implementation of the OAM sorter, the microscope is operated in scanning TEM (STEM) mode, with a small electron probe formed in the sample plane (Fig. 1a). The electron beam is then diffracted by the objective lens (Fig. 1b) into a disk in its back focal plane, where the first element of the sorter is located (Fig. 1c). The phase shift associated with the first element (S1) can be described mathematically by the expression

$$\varphi_{S1}(x,y) = \frac{ks}{f}\left(y \tan^{-1}\frac{y}{x} - x \log\left(\frac{\sqrt{x^2+y^2}}{L}\right) + x\right), \quad (1)$$

where $k$ is the electron wavevector, $f$ is the focal distance between the two sorter elements and $s$ and $L$ are scaling parameters. Element S1 can be fabricated using a long electrically charged conductive needle [22]. The scaling parameters $s$ and $L$ depend on the electrical bias applied to the needle and on its length, respectively [41]. Element S1 introduces a conformal mapping from Cartesian to log-polar coordinates (Figs 1c, d), which is concluded in the selected area aperture plane (Fig. 1f). Element S2 is located in this plane, which is conjugate to the sample plane. In its experimental realization, it comprises a periodic array of conductive needles, which are charged positively and negatively in an alternate fashion. The phase shift introduced by element S2 can be written in the form

$$\varphi_{S2}(u,v) = -\frac{ksL}{f} e^{-\frac{u}{s}} \cos\left(\frac{v}{s}\right), \quad (2)$$

where $u$ and $v$ are coordinates in the diffraction plane. This phase shift is designed to match and compensate (Fig. 1g) the phase shift of the electron beam, in order to prevent further S1-based evolution of the electron beam shape upon propagation. A second lens (Fig. 1h) collapses the electron beam to a rod (Fig. 1i), whose position with respect to the optical axis is proportional to the OAM carried by the electron beam. If several OAM components are present, then each component is focused to a different position, thereby generating a spectrum.

As a result of the stationary phase condition, the phase shifts of elements S1 and S2 vary rapidly. In the S2 plane, a phase change of $\Delta\phi=2\pi$ typically occurs over less than 200 nm (in the conditions used in the present work). Therefore, precise alignment of the elements is important to obtain complete phase compensation. This compensation involves controlling rotation, size, focus and translation.

Figure 2 shows the influence of different parameters on the resolution of an OAM spectrum. The three columns show electron optical simulations of the primary misalignments between the diffraction of S1 (left column) and S2 (centre column) and their effect on an OAM spectrum (right column). The simulations were performed for a 300 kV electron beam defined by a 2 mrad hard aperture traveling in vacuum. The sorter parameters are $s = 3$ μrad and $L = 40$ μm. The algorithm that was used for the calculations is described below. The first row corresponds to the ideal situation, in which the phase of the beam is matched perfectly and compensated by the phase of element S2 (centre column). The OAM spectrum then only features a narrow spot at $\ell = 0$ (right column). In the second row (Fig. 2b), the defocus of the objective lens (*df*) is considered. The conformal mapping is then not completed in the plane in which element S2 is located. As a result, a residual curvature is present in the beam and the phase pattern is distorted. The main effect of this aberration is broadening of the beam in the radial direction ($P_r$), which depends on the sign of the defocus. The OAM resolution is preserved instead, meaning that a moderate amount of defocus $f$ can be tolerated if only the OAM component is of interest [42, 43, 44, 45, 46].

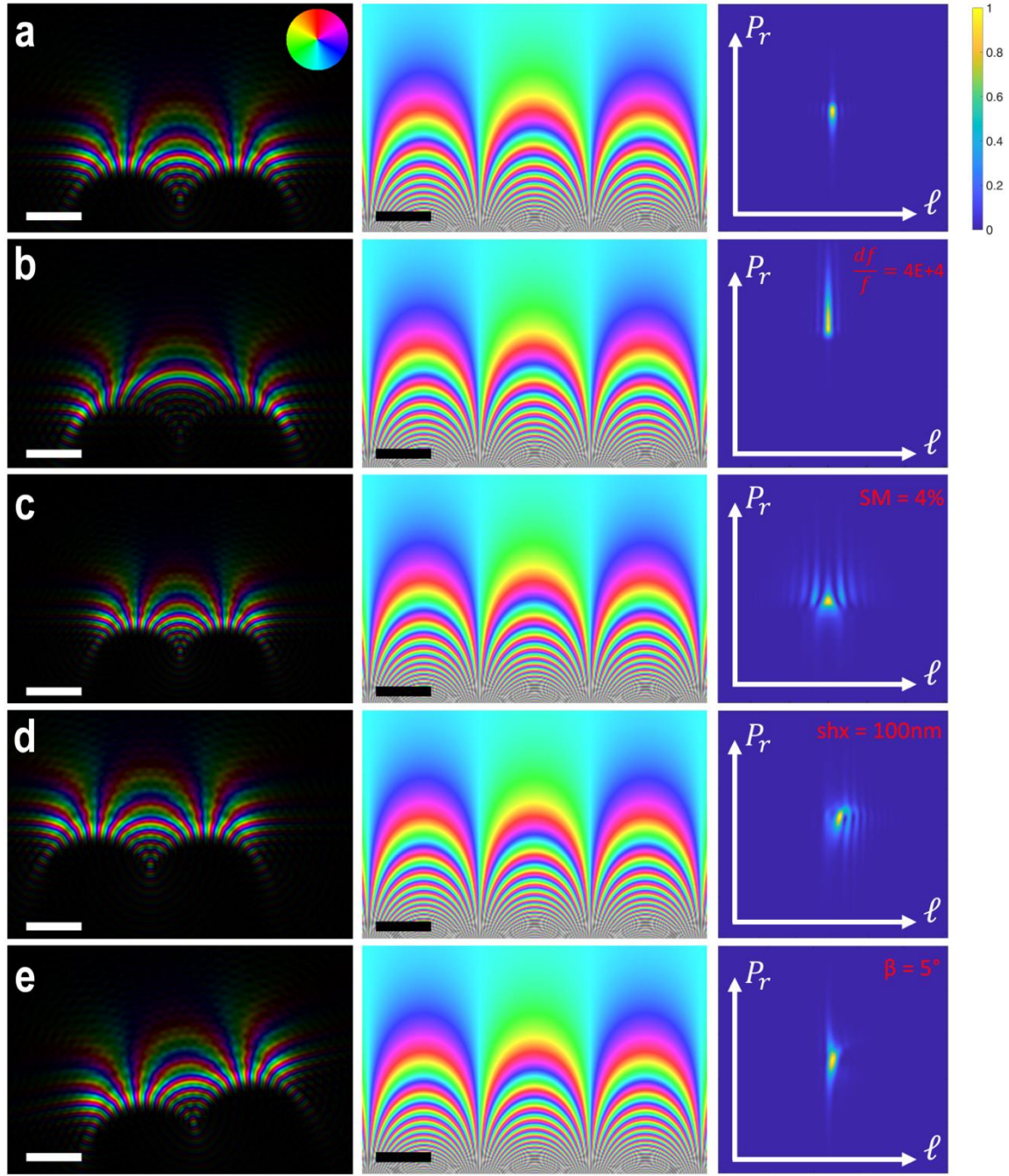

**Figure 2: Primary misalignment conditions between the diffraction of S1 (left column) and S2 (centre column) and their effects on an OAM spectrum (right column). a)** Perfect condition, with the beam and S2 aligned. **b)** Effect of objective lens defocus (*df*). **c)** Effect of size mismatch (*SM*). **d)** Effect of rigid shift (*shx* and *shy*). **e)** Effect of rotation (*β*). Scale bar in real space images: 10 μm. The OAM spectrum spans 40ℏ. The real space misalignment (left column) is magnified for better visualization. The real parameters that were used for the simulation are shown as labels.

A more prominent source of resolution loss is the size mismatch (SM) between the beam and element S2 (Fig. 2c). This aberration appears when the electron beam scale in the S2 plane is larger (smaller) than the scale dictated by the periodicity of element S2. Since the periodicity of S2 is bound to the physical distance between the electrodes and cannot be modified, correction for this aberration entails changing the beam size. The size of the electron beam in the S2 plane, in turn, depends on the product of the focal distance between the two planes (another physical parameter that cannot be changed) and the excitation parameter $s$ of element S1 (Eq. 1). An optimal potential then needs to be applied to the main electrode of element S1. A deviation from this value introduces an SM aberration to the OAM spectrum. The main effect of the SM aberration is the introduction of background fringes and a loss of OAM resolution. A difference of a few % is sufficient to produce the effect reported in the right column of Fig. 2c.

The third effect is associated with a rigid shift of the electron beam in one or both directions (*shx* and *shy*) with respect to element S2. The primary effect of this misalignment is an asymmetrical background fringe pattern. A misalignment of a few nm, which is small compared to the periodicity of element S2 (typically 20 µm), is sufficient to reduce the OAM resolution. Fortunately, TEMs offer high precision control over the beam position and this aberration can be fixed easily, even manually.

The last row reports the effect of relative rotation (*i.e.*, orientation mismatch $\beta$) of the electron beam with respect to element S2.

In addition to these effects, which are inherent to S1/S2 alignment, the defocus of the diffraction lens located after element S2 was also considered. Its effect is to modify the shape of the OAM spectrum. All other aberrations of the microscope, in particular astigmatism in different planes and coma, were neglected. At this stage, we assume that precise alignment is carried out before starting an experiment and it is not affected by a change of the sorter parameters. Aberrations of the main lenses can be included in the training of the CNN by adding more adjustable parameters. However, special care is devoted to the finite lateral coherence of the beam, as described below.

Even for a perfectly aligned OAM sorter, other issues can affect final experimental spectra. In particular, if the electron beam is not centred with respect to the optical axis in the sample plane or S1 plane, the OAM spectrum will be broadened. This is a normally unwanted effect that arises from the physical definition of OAM. The OAM operator, like its classical counterpart, is related to the specific choice of a pole, which corresponds here to the optical axis.

- **CNN and training dataset**

Figure 3 shows the structure of the CNN that was optimized here. It is composed of 5 convolution layer filters, each of which is followed by an average-pooling layer filter. Two fully connected layers lead to the output, corresponding to a total of 2,480,326 trainable parameters. The chosen activation function was the rectified linear unit (ReLU). The learning algorithm used was Adam [47] and the learning rate was 0.001, while the loss function was the root mean square difference between the predicted misalignment coefficients and the true coefficients. The neural network was implemented using the Keras library [48] and the TensorFlow backend [49].

The CNN was trained on a dataset of 20,000 simulated images (+ 2000 images for validation) for random values of the six misalignment parameters.

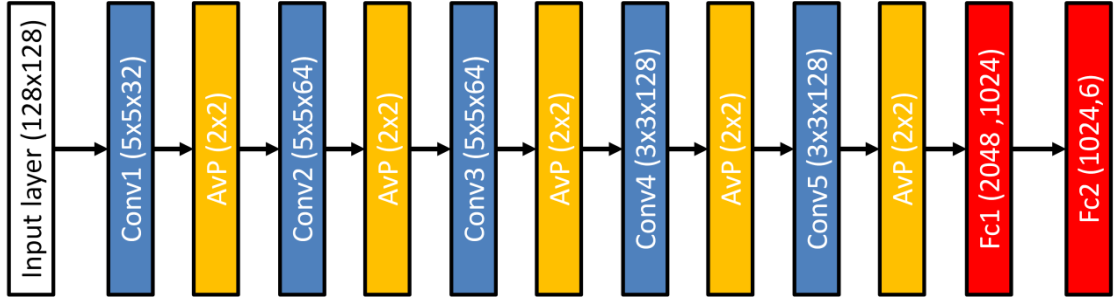

Figure 3: Structure of the optimized convolution neural network. Blue, yellow and red boxes correspond to convolution layers, average pooling layers and fully connected layers, respectively. Filter sizes are shown between parentheses.

Training can also be performed experimentally by changing the sorter parameters. Such a procedure is expected to improve the fitting because it will not be limited to an ideal numerical model. However, it requires complex automatic control of the microscope and long-term stability of microscope alignment. These problems are avoided here by training the CNN on simulated images.

In order to produce the training dataset, a numerical method was used to simulate the OAM spectroscopy experiment, based on a full wave calculation and free space propagation, in order to describe the effects of different aberrations and misalignments on the final resolution. The beam was free-space propagated between the elements using the Fresnel-Kirchhoff integral

$$U_z(u,v) = \frac{e^{ikz}}{i\lambda z} \iint U_0(x,y)\, e^{-ik\frac{xu+yv}{z}} dx dy \; , \qquad (3)$$

while the lenses were defined by quadratic phase elements of the form

$$T = exp\left(\frac{i(x^2+y^2)}{2f\lambda}\right) \; . \qquad (4)$$

The calculations were performed numerically on an 8k x 8k mesh using a Fourier transform algorithm for convolutions. Versions of the code were written in Matlab (for testing purposes) and C (for fast parallel computing). Entries from the training database are shown in Fig. 4 (top line).

Before being fed to the CNN, spectra were pre-processed to account for decoherence effects. Spatial and temporal coherence effects in electron microscopy are usually described by convolution and/or multiplication by suitable damping functions. A systematic treatment of coherence effects on the resolution of an OAM sorter is beyond the scope of the present paper and will be presented elsewhere. We accounted for decoherence using Monte Carlo simulations by averaging over OAM spectra obtained for different beam positions in the sample plane. The size of the deflection was evaluated by considering broadening of the probe between 0.5 and 1 Å. When magnification and propagation were accounted for, a best match was obtained for Gaussian broadening of 0.5ℏ in the OAM direction. Conversely, nearly no broadening appeared to be necessary in the radial direction. Without accounting for decoherence, the CNN did not converge to the correct value.

The CNN was trained on the simulation database for 20 epochs (with a computation time of 380 s per epoch), reaching a limit RMS error on the predicted parameter of 0.0038. Training was interrupted after 20 epochs, as the CNN started to show signs of overtraining, *i.e.*, the RMS error on the training dataset became lower than that estimated on the validation dataset, suggesting that the

CNN was memorizing features from the training images and losing its generalization ability. In order to validate the fitting accuracy of the CNN, the predicted misalignment coefficients were fed back into the simulation algorithm. The resulting images (Fig. 4, lower row) agree with the real images.

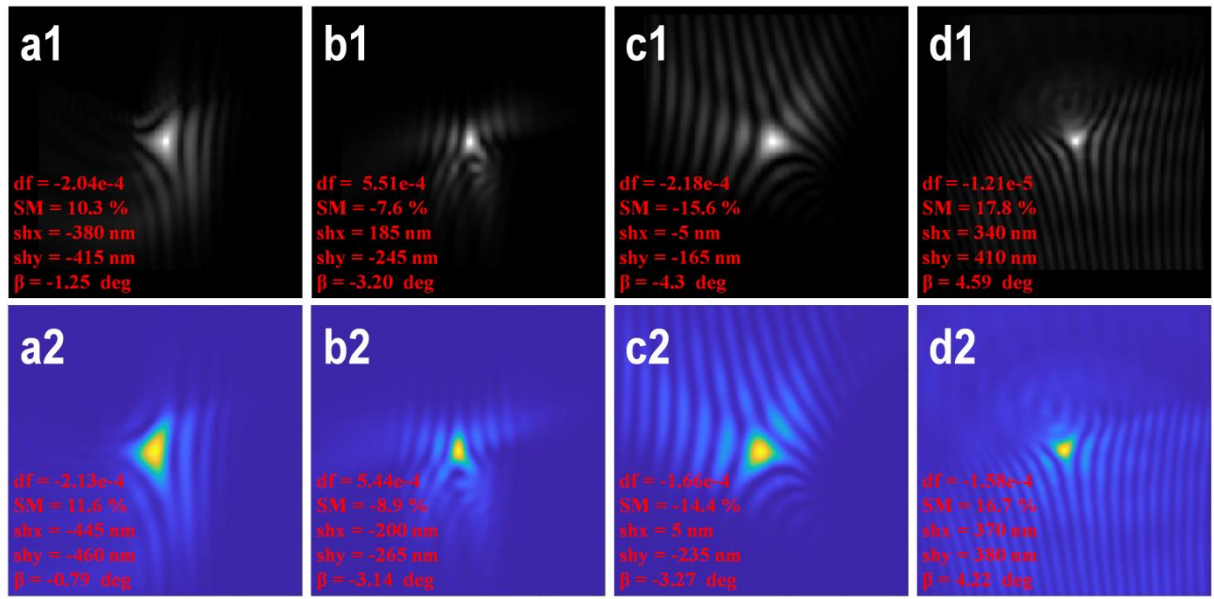

Figure 4: Demonstration of the predictive ability of the CNN for a comparison between simulated reference images (upper row) and simulations performed using the parameters predicted by the CNN (lower row).

**Experimental results**

Experiments were performed at 300 kV in a Thermo Fisher Titan G2 60-300 TEM equipped with an X-FEG emitter and an image spherical aberration corrector. The illumination system was set to spot size 9 and the three condenser lens system was used to achieve a probe convergence semi-angle of 2 mrad in the specimen plane. No sample was used, in order to allow free propagation of the electron beam in vacuum. The microscope was operated in "microprobe" mode with the objective lens at a standard pre-set value. The image aberration corrector was switched off, in order to achieve a larger focal distance between the sorting elements. In this configuration, element S2 was located in the diffraction plane of element S1 (mounted in the objective (OBJ) aperture plane), which is conjugate to the sample plane. The OAM spectrum could then be imaged on the detector with the microscope set to diffraction mode. The objective lens current was kept to a standard eucentric preset value, while focusing was achieved by changing the C3 excitation.

The device was first tuned to find a good working condition. The polarization of the main needle of element S1 was set to $V_c = 6.40$ V, corresponding to a scale factor $s \approx 3$ μrad. All experimental images were rescaled and rotated to match the scale and rotation of the training dataset before using the CNN. Calibration was achieved by sorting electron beams with known OAM signatures, as reported in [22]. Recorded OAM spectra were labelled based on the most sensitive parameters, i.e., the main potential applied to element S1 and the main defocus $f$. Mechanical instabilities of the aperture holders in which the sorter elements were mounted made precise calibration impossible.

The upper row of Fig. 5 shows experimental spectra recorded for different values of the potential applied to element S1 over a symmetrical range (6 - 6.8 V) about a reference value of 6.4 V, in order to study the influence of SM misalignment. The images were fitted using the CNN. The fitting parameters provided by the CNN were used to simulate spectra, in order to evaluate the qualities of the fits. The simulated spectra are shown in the lower row of Fig. 5. This comparison demonstrates the ability of the CNN to identify and analyse the main features of the spectra, in particular the arrowhead-like central peak, which points towards or away from element S2 (*i.e.*, up or down in the reference system used in Fig. 5), depending on the sign of SM.

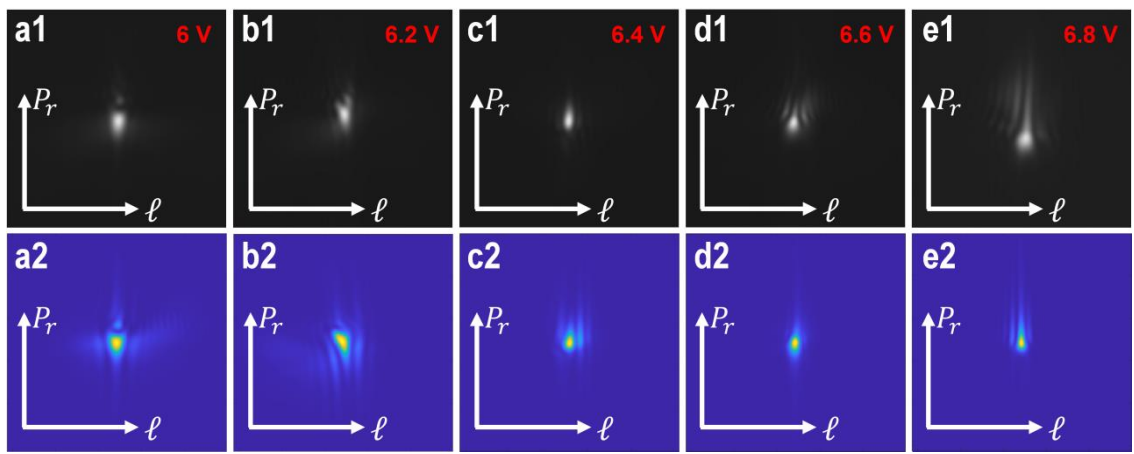

Figure 5: Upper row: Experimental spectra recorded for different values of the potential of element S1 over the symmetrical range (6 - 6.8 V). Lower row: Simulated spectra for best-fitting parameters obtained from the CNN.

The second parameter that was investigated was the defocus of the first lens (*df*). Spectra were again recorded over a symmetrical range about a reference value that provided the best resolution during manual alignment. The experimental results are reported in the upper row of Fig. 6, alongside fitting results in the lower row. The CNN is again able to reproduce the experimental images satisfactorily. The primary feature recognized by the CNN is streaking of the spectrum towards or away from element S2 (*i.e.*, up or down in the reference system used in Fig. 6), depending on the sign of the defocus.

Minor differences between the experimental and simulated spectra are thought to result from the algorithm used for simulation of the training dataset, which is based on an ideal sorter model. In reality, the phase of the sorter may be affected by imperfections in the shapes of the needles or contaminants on them. Conventional lens aberrations have also been neglected in the training model. A possible way to overcome these limitations would be to finalize the CNN training directly on experimental images after pre-training is performed on a simulated database.

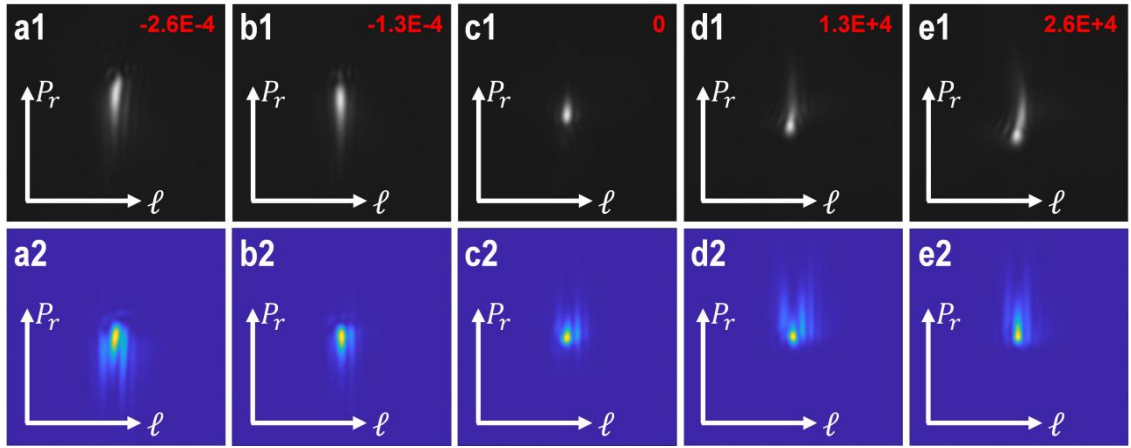

Figure 6. Upper row: Experimental spectra recorded for different values of the defocus of the objective lens (*df/f*). Lower row: Simulated spectra for best-fitting parameters obtained from the CNN.

Figure 7 shows an assessment of the accuracy of the quantification, in the form of a comparison between the estimated and experimental values. Figure 7a shows the expected linear trend between the estimated SM values and the real voltages applied to element S1. Figure 7b shows a comparison between the estimated defocus values and the real defocus values (Fig. 6), which were obtained by varying the current according to the expression

$$\Delta f = -2f \frac{\Delta I}{I} \ . \tag{5}$$

Although the linear trend again suggests successful fitting of the sorter misalignments, in practice misalignment effects tend to combine together, resulting in a complicated parametric dependence. The CNN was therefore allowed to extrapolate values of all of the parameters. In our best manual alignment (Figs 5c and 6c), the S1 potential was underestimated and the defocus was overestimated, illustrating how aberrations can partially compensate for each other, just as spherical aberration and defocus can partially compensate under Scherzer conditions for the electron probe. This compensation highlights the fact that perfect alignment would be difficult to achieve manually.

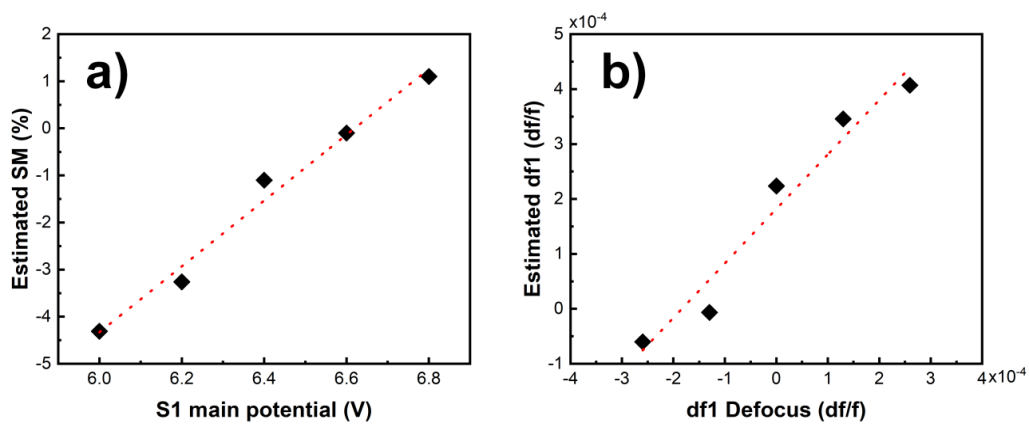

Figure 7: Comparison between experimental sorter parameters and CNN predictions shown as a function of a) S1 main potential and b) deocus (*df*).

The fitting of an experimental image using the CNN was found to only take (on average over 1000 iterations) 56 ms on a conventional laptop, including image pre-processing. The computational time is therefore negligible compared to the acquisition time. For this reason, we anticipate that a CNN can provide real-time control and feedback about alignment accuracy during experiments.

**Conclusions**

We have demonstrated that a neural network can be used to determine alignment parameters for the complex electron optical configuration of an OAM sorter, for which the effects of misalignment cannot easily be managed analytically or adjusted manually. The CNN is capable of determining parameters such as defocus and sorter electrode excitation from a single spectrum image. Such an approach can be applied in real time to align other complex optical systems, such as spherical aberration correctors, based on minimal experimental data. We envisage that in the future experimental devices will be able to self-diagnose and communicate with operators in real time.

**Acknowledgments**

The authors acknowledge the support of the European Union's Horizon 2020 Research and Innovation Programme under Grant Agreement No 766970 Q-SORT (H2020-FETOPEN-1-2016-2017).